# Parallel fast random bit generation based on spectrotemporally uncorrelated Brillouin random fiber lasing oscillation

Yuxi Pang [1,2], Shaonian Ma [1,2], Qiang Ji [1,2], Xian Zhao [1,2], Zengguang Qin [2,3], Zhaojun Liu [2,3], Ping Lu [4], Xiaoyi Bao [5], Yanping Xu [1,2,*]

[1] Center for Optics Research and Engineering, Shandong University, Qingdao 266237, China;
[2] Key Laboratory of Laser and Infrared System of the Ministry of Education, Shandong University, Qingdao 266237, China;
[3] School of Information Science and Engineering, Shandong University, Qingdao 266237, China
[4] National Research Council Canada, 100 Sussex Drive, Ottawa, ON K1A 0R6, Canada;
[5] Physics Department, University of Ottawa, 25 Templeton Street, Ottawa, ON K1N 6N5, Canada.
[*] Corresponding author: yanpingxu@sdu.edu.cn

**Abstract**: Correlations existing between spectral components in multi-wavelength lasers have been the key challenge that hinders these laser sources from being developed to chaotic comb entropy sources for parallel random bit generation. Herein, spectrotemporally uncorrelated multi-order Stokes/anti-Stokes emissions are achieved by cooperatively exploiting nonlinear optical processes including cascaded stimulated Brillouin scattering and quasi-phase-matched four-wave mixing in a Brillouin random fiber laser. Chaotic instabilities induced by random mode resonance are enhanced and disorderly redistributed among different lasing lines through complex nonlinear optical interactions, which comprehensively releases the inherent correlation among multiple Stokes/anti-Stokes emission lines, realizing a chaotic frequency comb with multiple spectrotemporally uncorrelated channels. Parallel fast random bit generation is fulfilled with 31 channels, single-channel bit rate of 35-Gbps and total bit rate of 1.085-Tbps. National Institute of Standards and Technology statistic tests verify the randomness of generated bit streams. This work, in a simple and efficient way, breaks the correlation barrier for utilizing multi-wavelength laser to achieve high-quality spectrotemporally uncorrelated chaotic laser source, opening new avenues for achieving greatly accelerated random bit generation through parallelization and potentially revolutionizing the current architecture of secure communication and high-performance computation.

## Introduction

Random bit generator (RBG) plays a vital role in various fields of scientific research, technology, and daily life, including cryptographic communications[1, 2], Monte-Carlo computational simulation[3], coordination in computer networks[4], lotteries[5] and stochastic modeling[6, 7]. Though it is easy to realize random bit generation in computers by pseudo-random bit generator with properties such as uniformity, independence, easy access and rapid generation, the reproducibility induced by the deterministic algorithms makes the generated random bits susceptible to prediction, thus deteriorating the information security of cryptographic communications[8]. Unlike the algorithmically generated pseudo-random bits, physical RBG is implemented through utilizing uncertainties of physical phenomena. Conventionally, physical RBGs are implemented by the chaotic electronic processes including frequency jitter of oscillators, thermal noise of circuits by using application-specific integrated circuit (ASIC) [9, 10] and field programmable gate array (FPGA)[11, 12]. However, these electronic RBGs suffer from low rates due to narrow bandwidths of the electronic chaotic sources. Over the past few years, various optical processes have been utilized to implement the physical RBGs, which successfully enhance the rate of generated random bit streams with improved randomness by exploiting the ultra-fast and chaotic optical nature. These optical RBGs have undergone great development with random bit generation of high rates realized through amplified spontaneous emission[13, 14], chaotic semiconductor lasers[15–17], single-photon radiation[18], quantum uncertainties[19–21], and spontaneous Raman scattering[22], etc.

Parallelization has been demonstrated as a simple and effective technique to enhance the generation rate and the scalability of RBGs by producing multiple parallel bit streams simultaneously. Parallel RBGs (PRBGs) are able to not only multiply the random bit throughput rate by combining their multiple-channel output, but also address the urgent demand of multiple data channels that need to be proceeded simultaneously in the current increasingly prevalent parallel and high-speed information architectures [23]. In the early reported works, physical PRBGs were implemented by methods based on spatially uncorrelated optical processes including two-dimensional imaging[24], incompatible spatial multiplexing[25, 26] and laser arrays[27, 28]. These schemes are basically inseparable from the massive processing element array, strict injection or feedback configurations, which complicate the operation process of PRBG. Moreover, channel scalability for these PRBGs is often at the cost of significant increases in complexity and cost. These limitations hinder the widespread practical deployment of PRBGs[29].

Recently, with the continuous breakthrough and application of optical wavelength/frequency division multiplexing technologies, realization of large-channel-capacity and parallel transmission of optical signal by means of wavelength multiplexing and demultiplexing becomes promising. In this context, multi-wavelength laser source shows a broad prospect as a promising candidate for constructing PRBGs through parallelization in wavelength or frequency domains. However, attempts on realization of PRBGs in wavelength/frequency domain based on multi-wavelength laser sources have encountered remarkable challenges due to the high correlations between different spectral components owing to the coherent coupling between longitudinal modes, making them single-channel [30] or incompletely uncorrelated multi-channel [29] devices. In the past few years, random fiber lasers utilizing Rayleigh scattering-based mirrorless feedback have sparked great interest owing to their simple design and unique properties. By taking advantage of their chaotic dynamic nature [31], single-wavelength random fiber lasers based on various gain mediums and random feedback mechanisms have been demonstrated as the physical entropy source for constructing single-channel RBG [31–34]. It is naturally implied that a multi-wavelength random fiber laser would possess the potential to break the correlation barrier that exists in conventional multi-wavelength laser sources to achieve RBG with multiple parallel channels. Although multi-wavelength random fiber lasers have been studied in several works [35–39], it is still unclear whether they can function as efficient parallel chaotic sources.

In this paper, we fill this gap by proposing a parallel RBG based on a spectrotemporally uncorrelated multi-wavelength random fiber laser. For the first time, to the best of our knowledge, we verified the inter/intra-channel spectrotemporally uncorrelation between multi-wavelength lasing oscillations in a random resonant cavity, proving the capability of parallelization. A proof-of-concept PRBG based on a spectrotemporally uncorrelated Brillouin random fiber laser (STU-BRFL) is demonstrated to implement the spectral parallelization for random bit generation. With the combined implementation of the cascaded stimulated Brillouin scattering (CSBS) and four-wave mixing (FWM) processes, a multi-wavelength random lasing oscillation with up to 15 orders of Stokes lights and 15 orders of anti-Stokes lights is achieved. Benefitting from the chaotic instabilities induced by the random distributed feedback and intense mode competition that is enhanced and disorderly redistributed among multiple Stokes/anti-Stokes lights through the CSBS process and the quasi-phase-matched FWM process, the inherent correlation between multiple Stokes and anti-Stokes lights is effectively released and maintains at a low level. Meanwhile, a logarithm-based high-order derivative algorithm is employed on the digitized random lasing intensity of the STU-BRFL to generate random bit sequences by retaining a number of the least significant bits (LSB) of the high derivative value, eventually realizing a 31-channel PRBG with rates up to 35-Gbps for single-channel operation and 1.085-Tbps for combined-multi-channel operation by using the first 15 orders of Stokes lights, the first 15 orders of anti-Stokes lights and the residual pump light. The randomness of the generated bit streams is verified by National Institute of Standards and Technology (NIST) statistics tests. In addition, the proposed STU-BRFL-based PRBG also features the convenient channel scalable capability, which is crucial for boosting the bit rate and channel number of PRBG.

# Results

## 2.1 Principle of operation

Fig. 1 shows the schematic for the STU-BRFL-based PRBG. The operation principle of STU-BRFL is shown in Fig. 1(a). Pump light injected into the Brillouin gain fiber triggers the SBS process along the fiber, generating and amplifying the backward-propagating Stokes light through the electrostriction process. Then the generated Stokes light is guided into the Rayleigh scattering fiber, which is subsequently randomly backscattered along the fiber through Rayleigh scattering. The backscattered Stokes light then re-enters the Brillouin gain fiber and rejoins the Brillouin amplification process as the seed. Once the accumulated Brillouin gain overcomes the total optical loss, the Stokes light is able to resonate in the random cavity. As shown in Fig. 1(a), by re-injecting the backward-propagating resonant Stokes lights into the Brillouin gain fiber in the same propagation direction with the pump light, the CSBS process can be induced. In the CSBS process, low-order Stokes lights serve as the pump lights for generating neighboring higher-order Stokes lights, eventually leading to the multi-wavelength random lasing oscillations as shown in Fig. 1(b). Note that the anti-Stokes light of each order is also generated and propagates along with the Stokes lights during the CSBS process, which however is very weak in the absence of other nonlinear amplification processes. Here, this absence of nonlinear amplification can be well compensated by the occurrence of FWM process. In the Brillouin gain fiber, once phase matching condition for the pump light and the generated Stokes/anti-Stokes lights are satisfied with proper management on dispersions and nonlinearities, the FWM process would occur among lights in both directions in the main cavity, redistributing optical powers among the pump light and the Stokes/anti-Stokes lights. This process boosts the power of existing Stokes lights and facilitates the generation of higher-order Stokes and anti-Stokes lights. The bottom part of Fig. 1(b) shows the FWM process occurring among the pump light and Stokes/anti-Stokes lights which act as the pump or signal light (marked by the red triangles) to give rise to the idler lights (marked by the blue triangles). In fact, all the pump light, Stokes lights in the CSBS process and the newly generated anti-Stokes lights during the FWM process can act as the pump or signal light to participate in the FWM process to produce other idler light, as illustrated by the six FWM processes with different combinations of pump, signal and idler lights in Fig. 1(b). It is worth noting that a phase-matching condition is regarded to be strictly met for the FWM process in the case when the phase mismatch factor is equal to zero, which however is difficult to be achieved in most cases. In the current scenario, the FWM process occurs with quasi-phase-matching condition, i.e., the phase mismatch factor is not equal to but close to zero, in which the efficiency or the parametric gain of the FWM is not optimum. Thus, the FWM process can effectively take place for light waves with a certain bandwidth, as long as phase mismatch factor is close to zero for any spectral component within the associated light waves. With the cooperative interaction between the CSBS process and the FWM process, a large number of higher-order Stokes and anti-Stokes lights can be effectively generated and enhanced in the laser cavity, facilitating the random laser comb generation with multi-orders of Stokes and anti-Stokes lights.

In each emission line of the STU-BRFL, resonant lasing spikes appear at stochastic frequencies and keep drifting on top of the Brillouin gain spectrum owing to the thermally induced instability in Rayleigh scattering as well as the intense mode competition. The presence of Rayleigh scattering centers randomly distributed along the Rayleigh fiber contributes to numerous densely-spaced longitudinal modes with random spacing in the resonant cavity, as shown in the inset of Fig. 1(b). The amplitude of the resonant spectral components within the Brillouin gain spectrum fluctuates with time, which is induced by the intense mode competition within the inhomogeneous broadening Brillouin gain spectrum. The randomly distributed Rayleigh scattering centers introduce stochastic delays for the randomly-spaced resonant modes, which significantly decreases the phase correlation between the intra- or inter-line random modes and further enhances the intensity fluctuations of the resonant light. It should be noted that all the Stokes/anti-Stokes lights are generated and amplified through several nonlinear optical processes, including the CSBS process, FWM process as well as the possible modulation instability process that require phase matching condition for occurrence, which would inevitably introduce inter-line correlations between the multiple emission lines. However, the random mode resonance process effectively prevents the construction of these correlations. The resonant frequency of random modes within the Brillouin gain bandwidth of each order is determined by the locations of the Rayleigh scattering centers which are randomly distributed along the Rayleigh scattering fiber and randomly changed in each round-trip of the resonant light within the open cavity. Benefiting from the wavelength-independent nature of this disordered process, the resonant random modes within Brillouin gain spectrum of different orders are independent with each other. Furthermore, mode competitions among the densely-spaced random modes take places locally and independently for Stokes/anti-Stokes lights of different orders, leading to uncorrelated amplitude fluctuations in Stokes/anti-Stokes emissions of different orders. Thus, the inherent dependence in the CSBS process is effectively eliminated by the random mode resonance. Moreover, the FWM process redistributed the energy among the pump, signal and idler lights, introducing the inter-line correlation in the multi-wavelength random lasing, as illustrated in the bottom part of Fig. 1(b). However, in the quasi-phase-matched condition, the densely-spaced random modes in each order Stokes/anti-Stokes light participate in the FWM process as the pump or signal modes to generate the idler modes, resulting in the random combination and overlapping of idler modes within the Brillouin gain bandwidth of each order. In addition, in the quasi-phase-matched condition, FWM process is more susceptible to noises and more easily to be disturbed while providing the parametric gain to Stokes/anti-Stokes lights, which thus introduces intense instabilities into the generated idler lights. Hence, the inter-line correlation induced by the FWM process is also effectively eliminated. In addition, modulation instability would also occur when the wavelengths of the resonant lights fall in the anomalous dispersion regime of the Brillouin gain fiber with a high nonlinear coefficient, which leads to the exponential growth of co-propagation noise due to the interplay between the nonlinear and dispersion effects and thus imposes random temporal intensity modulations on the resonant Stokes/anti-Stokes lights, further enhancing the temporal randomness of the laser output along with the random mode resonance. Ultimately, the temporal and spectral correlations for the Stokes/anti-Stokes emissions are greatly minimized by the random mode resonance process and its interplay with the CSBS process and the quasi-phase-matched FWM process, giving rise to the spectrotemporally uncorrelated multi-wavelength random lasing oscillation which can be effectively utilized to work as a parallel optical chaotic source. In the experiment, the spectrotemporally uncorrelated random lasing oscillation is implemented by a dual-cavity BRFL, in which a 390-m-long highly nonlinear fiber (HNLF) provides the Brillouin gain and parametric gain and a 5-km-long single-mode fiber (SMF) provides the random distributed feedback. The experimental setup is sketched in Fig.2 and detailed in Methods: Experimental details of the STU-BRFL.

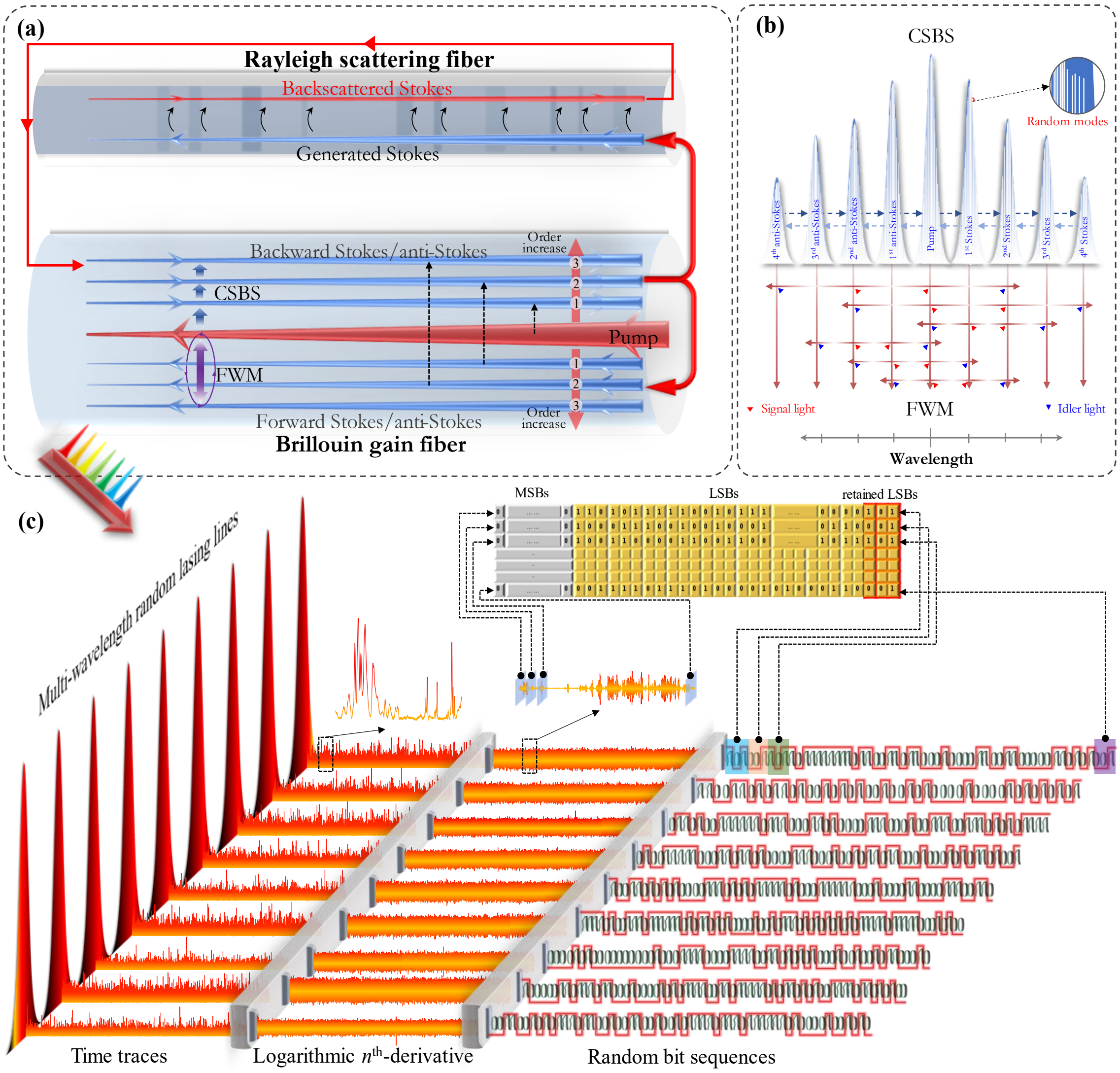

Fig. 1 | Schematic for the STU-BRFL-based parallel random bit generation. (a) Operation principle of STU-BRFL involving the CSBS and the FWM processes. (b) Mixed optical interactions including the CSBS, the FWM and the random mode resonance processes. (c) Flow chart for the generation of parallel random bit sequences.

The parallel random bit streams can be generated following the procedures illustrated in Fig. 1(c). Temporal output corresponding to each spectral component in the STU-BRFL is individually recorded after the random lasing emission is demultiplexed by a wavelength demultiplex device. The optical temporal trace of the selected Stokes/anti-Stokes line is then converted to electrical signal by a photodetector (PD). In the subsequent data processing, the method of logarithm-based $n^{th}$-derivative is proposed and used to eliminate the statistical bias of the random intensity fluctuation in Stokes/anti-Stokes lights. The distribution of results for the derivative function is highly symmetric, smooth and can be evenly divided into equal numbers of ones and zeroes by a mapping of even/odd bins to logical 0/1 values. Then, the *m*-LSB of the result of the logarithmic $n^{th}$-derivative are appended to the bit sequence to avoid the periodicity for large derivative values and meanwhile further increase the random bit generation rate. Finally, the randomness of the generated digital signals is tested by the NIST test suites.

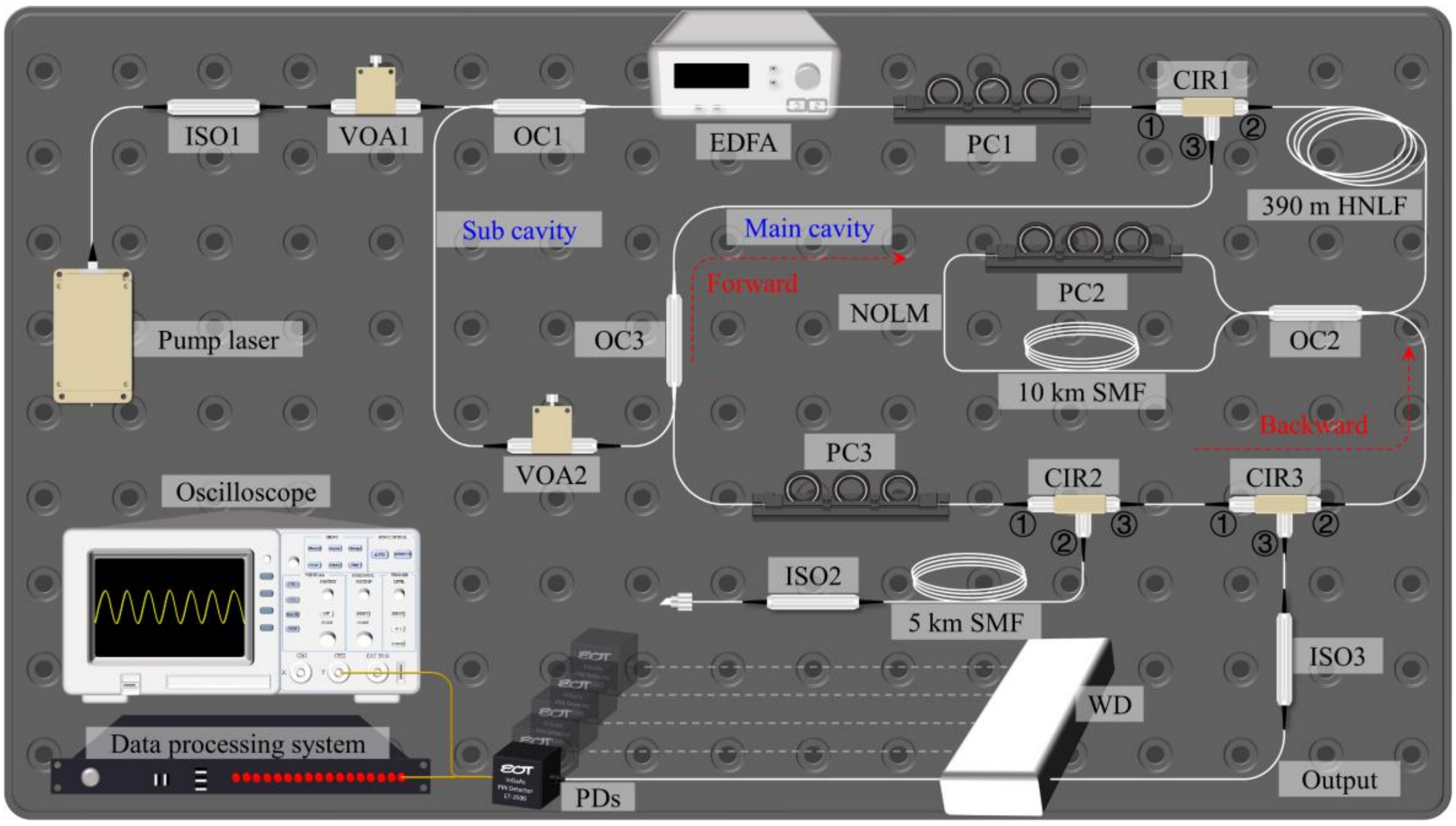

Fig. 2 | Experimental setup of the STU-BRFL-based PRBG. (ISO: isolator, VOA: variable optical attenuator, OC: optical coupler (OC1 and OC3: 1 × 2 50:50, OC2: 2 × 2 70:30), WD: wavelength demultiplexer, EDFA: erbium-doped fiber amplifier, PC: polarization controller, CIR: circulator, HNLF: highly nonlinear fiber, SMF: single-mode fiber, NOLM: nonlinear optical loop mirror, PDs: photodetectors.)

### 2.2 Random lasing oscillation of STU-BRFL

In the experiment, the output power of the semiconductor laser is boosted by the EDFA to 500 mW. The optical spectra of the STU-BRFL are monitored by an optical spectrum analyzer (YOKOGAWA, AQ6370D) and the results are shown in Fig. 3(a). It can be found that 15 orders of Stokes lights and 15 orders of anti-Stokes lights are generated through the CSBS process and the FWM process. Then, the output of each demultiplexed channel of WD is acquired by PDs and monitored by an oscilloscope. The time-domain traces of the 1st-order and 2nd-order Stokes lights, the residual pump light, and the 1st-order and 2nd-order anti-Stokes lights in a time span of 1s are plotted in Fig. 3(b1-b5), accompanied by their statistical distributions of amplitude in Fig. 3(c1-c5). Obviously, these time-domain traces exhibit a strong temporally chaotic behavior. The histograms of the recorded time-domain traces exhibit an asymmetric distribution with the counted spikes numbers following an exponential decrease. This is due to the fact that complex nonlinear optical interactions between the CSBS, the quasi-phase-matched FWM, the random mode resonance process as well as the possible noise-driven modulation instability process occur in the STU-BRFL, which significantly amplifies the chaotic instabilities induced by the random distributed feedback and intense mode competition and redistributes these chaotic instabilities among the resonant Stokes/anti-Stokes lights. Consequently, the continuous laser output is transformed into random self-pulsing one with extreme events captured in the amplitude distribution in Fig. 3(c1-c5), in which long tail is observed at large amplitude. In addition, the optical spectra and power spectra of demultiplexed Stokes and anti-Stokes lights are plotted and analyzed in Supplementary Note I.

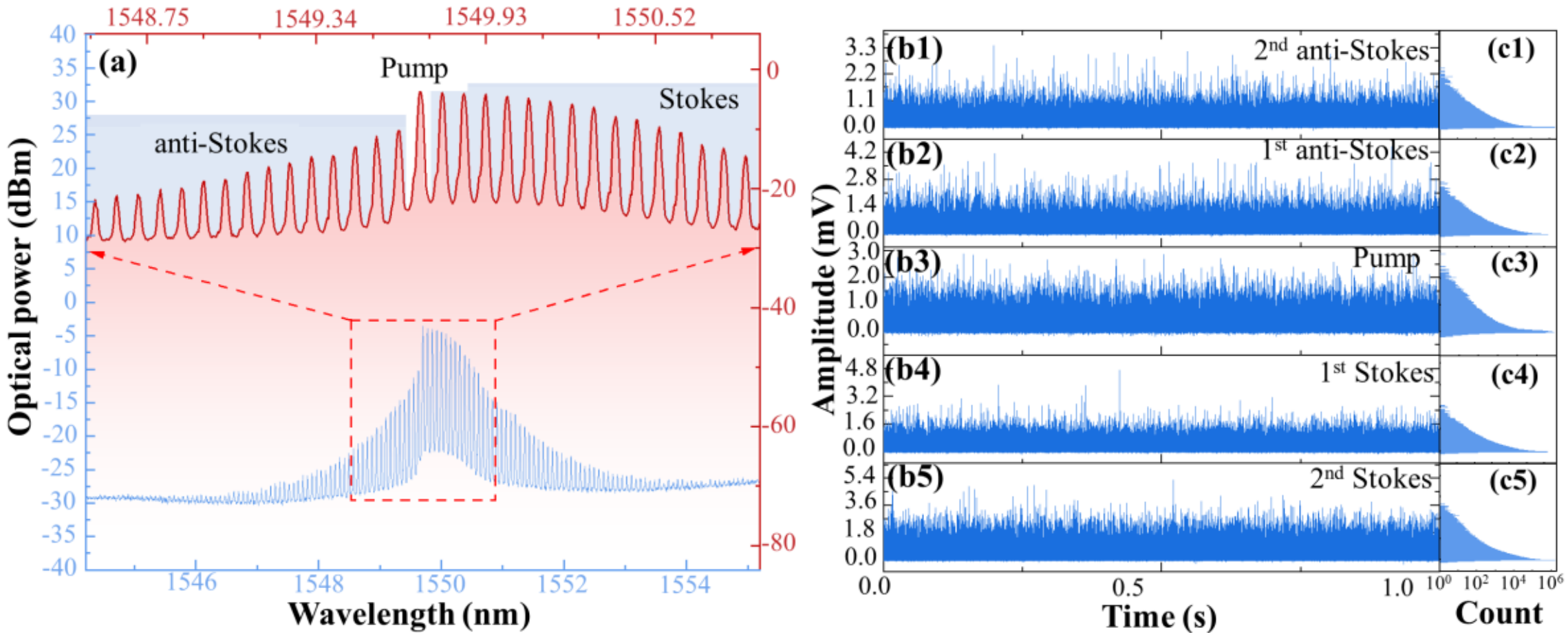

Fig. 3 | (a) Optical spectra of the STU-BRFL output; (b1-b5) Time-domain traces of 2nd-order anti-Stokes, 1st-order anti-Stokes, Pump, 1st-order Stokes, 2nd-order Stokes lights, and their statistical distribution of amplitude (c1-c5).

The intra-channel correlation calculated by auto-correlation function (ACF) [31] for the selected Stokes and anti-Stokes light as well as the residual pump light with their corresponding optical spectrum are shown in Fig. 4. Fig. 4(a1~a5) depicts the ACF results of the temporal traces extracted from 13$^{th}$-order and 1$^{st}$-order anti-Stokes lights, pump light, 1$^{st}$-order and 15$^{th}$-order Stokes lights as examples. Fig. 4(b1~b5) depicts the spectrum of the multi-wavelength output and the selected individually filtered light components. Obviously, all the ACF results present a $\delta$-function shape without any side peaks, indicating that no time delay signature exists in the random lasing output. Thus, unpredictability and randomness are ensured, demonstrating that each lasing line of the STU-BRFL can individually function as an effective chaotic light source.

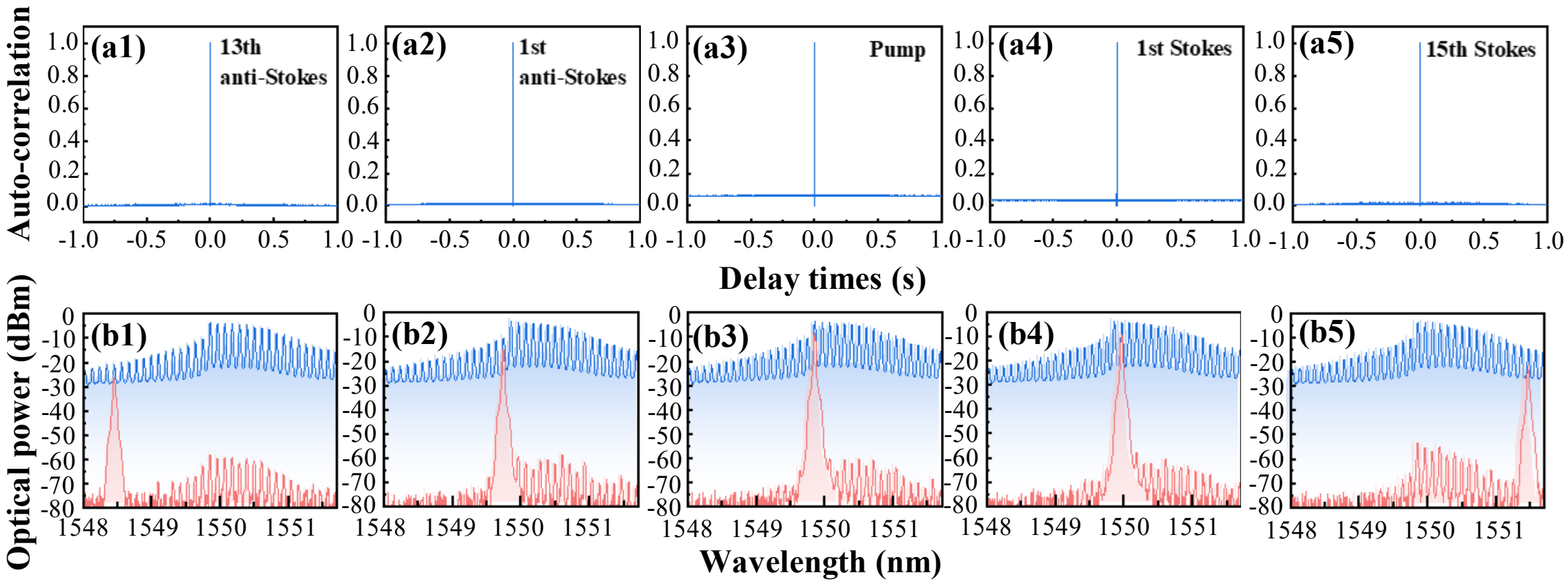

Fig. 4 | Auto-correlation functions of time traces acquired from 13$^{th}$-order (a1) and 1$^{st}$-order (a2) anti-Stokes, pump (a3), and 1$^{st}$-order (a4) and 15$^{th}$-order (a5) Stokes lights, and the corresponding optical spectrum of these lights (b1~b5).

The inter-channel correlation between the demultiplexed light components in the STU-BRFL is calculated by cross-correlation function (XCF) [30]. The calculated XCF results between 1$^{st}$-order and 5$^{th}$-order anti-Stokes lights, pump and 1$^{st}$-order anti-Stokes lights, 5$^{th}$-order Stokes and 5$^{th}$-order anti-Stokes lights, pump and 1$^{st}$-order Stokes lights, 1$^{st}$-order and 5$^{th}$-order Stokes lights are shown in Fig. 5(a1-a5) as examples. It can be found that none of the cross-correlation results exhibit any significant correlation peak and the calculated XCF coefficient level maintains near zero, which verifies the independence among the selected channels of the STU-BRFL output. XCFs are also calculated for Stokes/anti-Stokes lights and residual pump light with 1-, 2- and 3-order difference, with the evolution of the positive/negative maximum cross-correlation coefficients shown in Fig. 5(b~d), respectively. Obviously, all the results show that the calculated positive/negative maximum cross-correlation coefficients are close to zero, demonstrating that spectral correlations in these channel combinations are low, which satisfies the requirements for parallel random number generation. Fig. 5(b1) and (b2) represent the trends of the positive and negative maximum cross-correlation coefficient between neighboring spectral components of the multi-wavelength laser output. It can be seen that the adjacent anti-Stokes light above the 12$^{th}$ order exhibits a higher positive cross-correlation coefficient, while lower positive cross-correlation coefficients are exhibited for all the remaining neighboring spectral components. In Fig. 5(b2), the adjacent anti-Stokes lights from the 3$^{rd}$ to 11$^{th}$ orders exhibit lower negative cross-correlation coefficients, while the remaining spectral components exhibit higher coefficients. The gray shaded region in the figure indicates the region of spectral components with both small positive/negative cross-correlation coefficients, which mainly include the adjacent anti-Stokes lights of the 3$^{rd}$ to the 11$^{th}$ order. The main reason for this phenomenon is that the Stokes light and the lower-order anti-Stokes light are mainly generated or amplified by the CSBS process, and the phase-matched energy transfer between neighboring orders makes the adjacent spectral components exhibit relatively higher cross-correlation, while the higher-order anti-Stokes are basically generated by the quasi-phase-matched FWM process, in which phase matching condition is not as strictly met as in the CSBS process and more intense instability is introduced through the disordered energy redistribution, thus leading to the lower correlations between adjacent high-order anti-Stokes lights. Fig. 5(c) and (d) illustrate the evolutions of the positive/negative maximum cross-correlation coefficients of the Stokes/anti-Stokes lights and residual pump light with 2- and 3-order difference. Similar to the results exhibited in Fig. 5(b), most of the anti-Stokes lights exhibit weak dependence between spectral components, whereas the Stokes and the lower-order anti-Stokes lights show relatively higher cross-correlation coefficients. In addition, it can be clearly observed that the cross-correlation coefficient for the Stokes lights decreases with the order, which is mainly due to the fact that the pump light delivered order by order during the CSBS process gradually fails to provide sufficient Brillouin gain for the generation and resonance of the higher order Stokes lights, which are generated and amplified dominantly through the quasi-phase-matched FWM process. For an overall comprehensive characterization of the inter/intra-channel correlation of the STU-BRFL, the maximum absolute value of the correlation of each pair of 31 Stokes/anti-Stokes/pump lights is calculated and shown in Fig. 5(e). A significant line with relatively strong correlations can be observed at the main diagonal, which is the autocorrelation. Except for this line, low correlations are exhibited between the multiple Stokes/anti-Stokes/pump lines. Thus, the above ACF and XCF calculation results prove that output from the STU-BRFL strictly satisfies the requirement for constructing

PRBG. More details for the calculation of ACF and XCF are presented in Methods: Characterization of random laser output, and the full calculation results of ACF and XCF are depicted in Supplementary Note II.

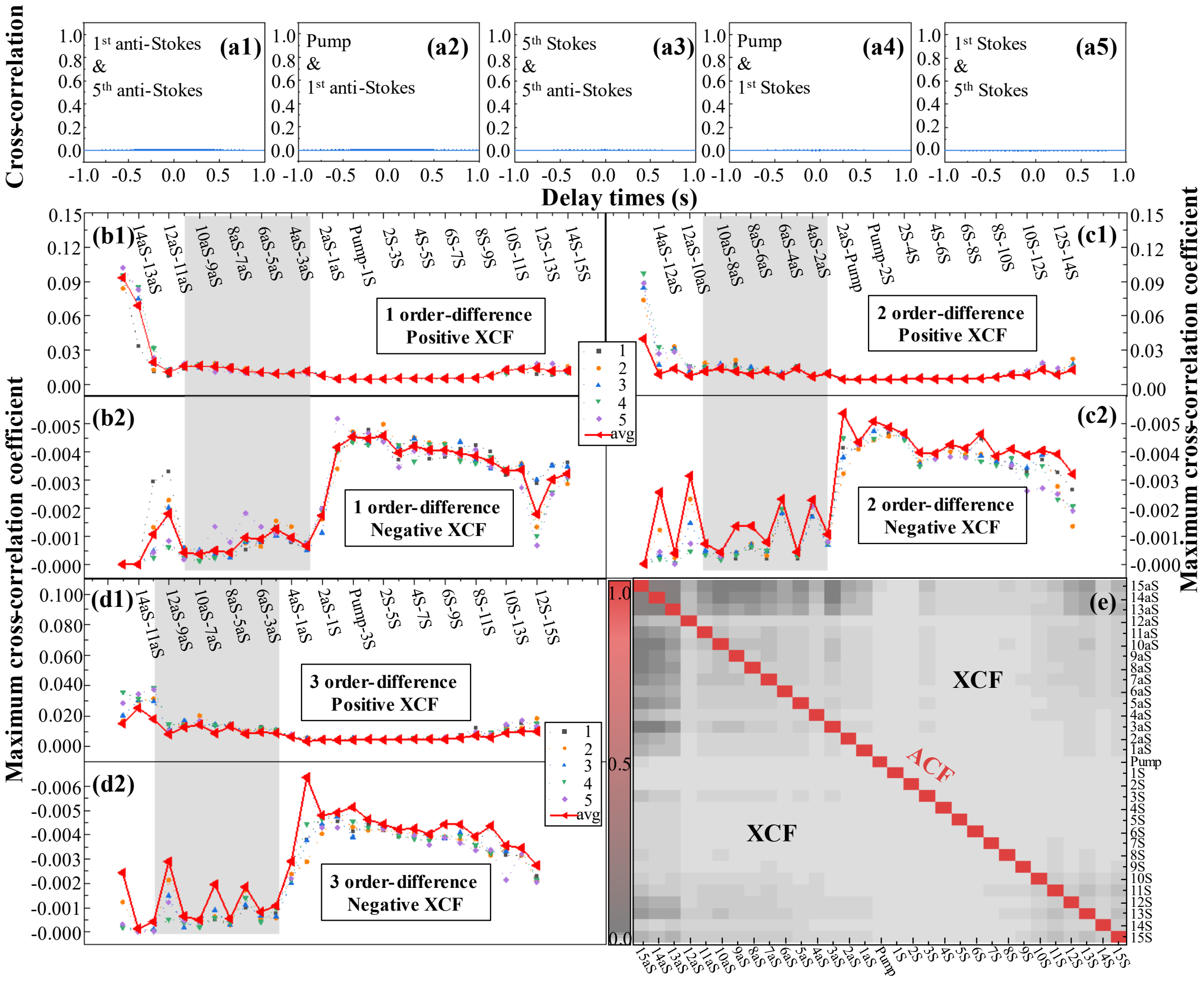

Fig. 5 | Cross-correlation function of time traces between 1st-order and 5th-order anti-Stokes lights (a1), pump and 1st-order anti-Stokes lights (a2), 5th-order Stokes and 5th-order anti-Stokes lights (a3), pump and 1st-order Stokes lights (a4), 1st-order and 5th-order Stokes lights (a5); and the evolution of the positive/negative maximum cross-correlation coefficient for Stokes/anti-Stokes lights as well as pump light with 1-(b1~b2), 2-(c1~c2) and 3-(d1~d2) order difference; (e) The maximum absolute value of the correlation between 31 different Stokes/anti-Stokes/pump lights.

### 2.3 Parallel random bit generation

The generation of the random bit stream in each channel consists of the following three steps. First, the logarithmic data processing is performed on the recorded chaotic signal, which can improve the randomness and the symmetry of the amplitude distribution of the chaotic signal. Second, the $n^{th}$ derivative is calculated by using $n$+1 successive values of the logarithmic-processed data, which not only improves the symmetry of the amplitude distribution but also optimizes the uniformity of the chaotic signal. Third, the $m$-LSBs of the results of the $n^{th}$ derivative are appended to the bit sequence to dramatically increase the random bit generation rate. The chaotic temporal trace of the random lasing oscillation in our experiments is first digitized at a 2.5 GS/s sampling rate with a 12-bit vertical resolution, and examples for the recorded $2^{nd}$-order Stokes light are shown in Fig. 6. Fig. 6(a1-a2) and Fig. 6(b1-b2) show the originally recorded sampling data and the logarithmic-processed data of the chaotic lasing output, respectively. In Fig. 6(a1-a2), the amplitude bias of the recorded chaotic signals is clearly observed, which results in an uneven distribution of the generated random bits with zeros and ones. This is mainly caused by the fact that complex nonlinear optical interactions between the CSBS, the quasi-phase-match FWM, the random mode resonance process as well as the possible noise-driven modulation instability process occur in the STU-BRFL, in which the chaotic instabilities induced by the random distributed feedback and intense mode competition is significantly amplified and redistributed among the resonant Stokes/anti-Stokes lights, results in the self-pulsing or even extreme events in the random lasing oscillation process. After the logarithmic procedure, the amplitude bias of the chaotic signal is reduced. Furthermore, the logarithmic-processed chaotic signal shows a more uniform amplitude distribution than the original data by comparing their time traces as shown in Fig. 6(a1) and Fig. 6(b1) and corresponding amplitude distributions as shown in Fig. 6(a3) and Fig. 6(b3), respectively. Since the $n^{th}$ derivative is extremely sensitive to small changes of the data in a time window of $n$+1 sampling points, using a sufficiently high derivative can effectively improve the randomness of the original data. The blue shaded region in Fig. 6(a2, b2) indicates the six successive sampling data points required for the calculation of the fifth derivative (red point in Fig. 6(a4, b4)), respectively. Fig. 6(a4~a5) and Fig. 6(b4~b5) show the calculated $5^{th}$ derivative of the original data and the logarithmic processed data in 0.04-μs and 5-ms time span, respectively. It is clear that the $5^{th}$ derivatives of both the original and the logarithmic-processed data exhibit favorable symmetry around zero and an obvious improvement in unbiasedness. The amplitude distribution of the $5^{th}$ derivative of the logarithmic chaotic signal exhibits better uniformity than that of the original sampling signal as shown in Fig. 6(a6, b6).

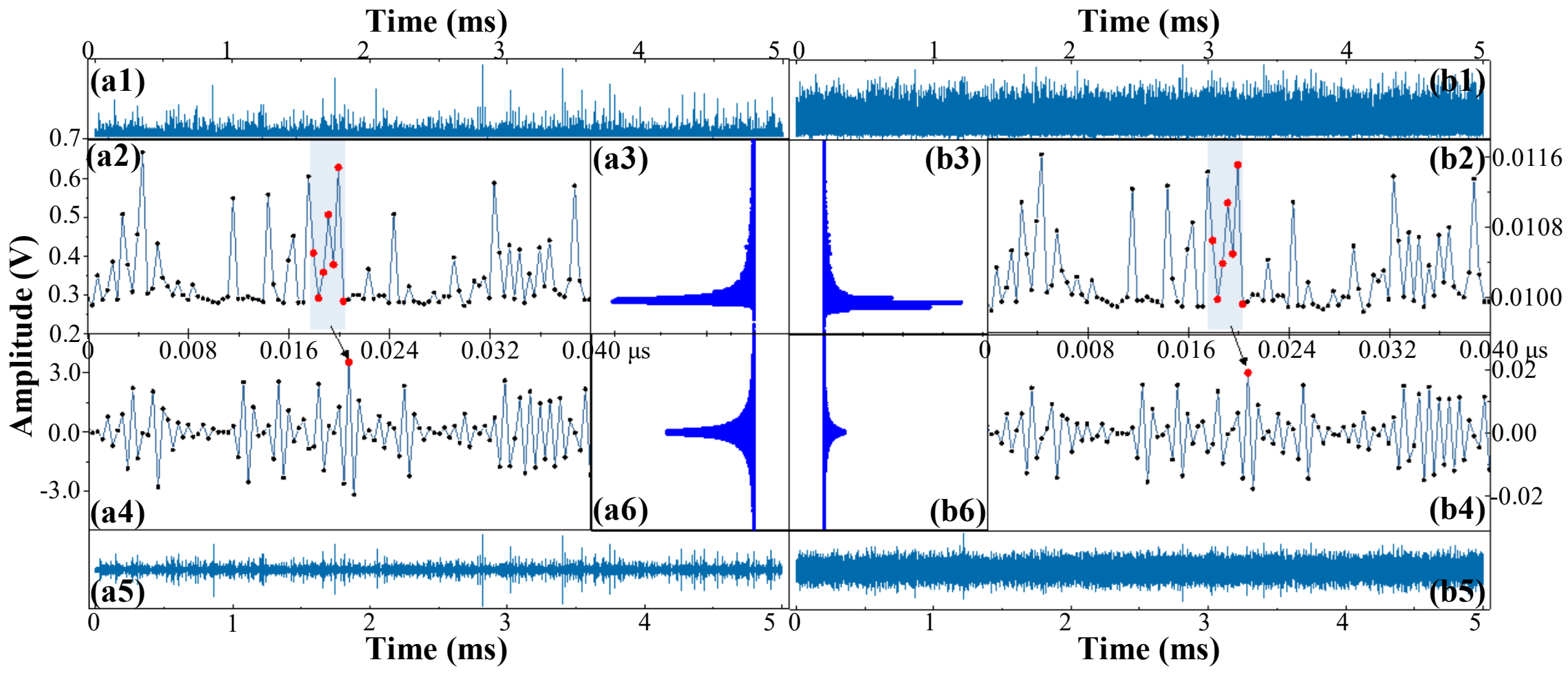

Fig. 6 | Examples of digitized data of 2nd-order Stokes light: Recorded original sampling data and the logarithmic processed data in 5-ms (a1, b1) and 0.04-μs (a2, b2) time span and their corresponding amplitude distribution (a3, b3), respectively. Fifth derivative of the original sampling data and the logarithmic processed data in 0.04-μs (a4, b4) and 5-ms (a5, b5) time span and their corresponding amplitude distribution (a6, b6), respectively.

The histogram of the logarithmic chaotic signal extracted from the 2nd-order Stokes light is shown by the solid red line in Fig. 7(a). This histogram shows an obvious asymmetric L-shaped distribution with severe irregularity and a significant deviation from the zero point. Besides the original logarithmic chaotic signal, the amplitude distribution of chaotic signals processed with derivative of different orders after the logarithmic process are also shown order by order in Fig. 7(a). It can be found that the chaotic signals processed by the $n^{th}$ derivative show an improved symmetric Gauss-like distribution centered around zero. Moreover, the amplitude distribution of chaotic signals becomes more consistent and uniform and covers a larger amplitude range as the order of the derivative process increases. This facilitates that more LSBs can be selected to be appended to the random bit sequences for achieving higher rates. At the same time, the logarithmic data processing also significantly increases the number of available LSBs by introducing a larger number of decimal places in the amplitude of the chaotic signal, which thus increases the rate and meanwhile enhances the randomness of the generated bit sequence. The digital post-processing procedure for random bit generation is illustrated in Fig. 7(b). The random bit sequences are generated at high rates by performing logarithmic and high-order derivative operations on original data, and the detailed data processing is described in the Methods: Random bit generation.

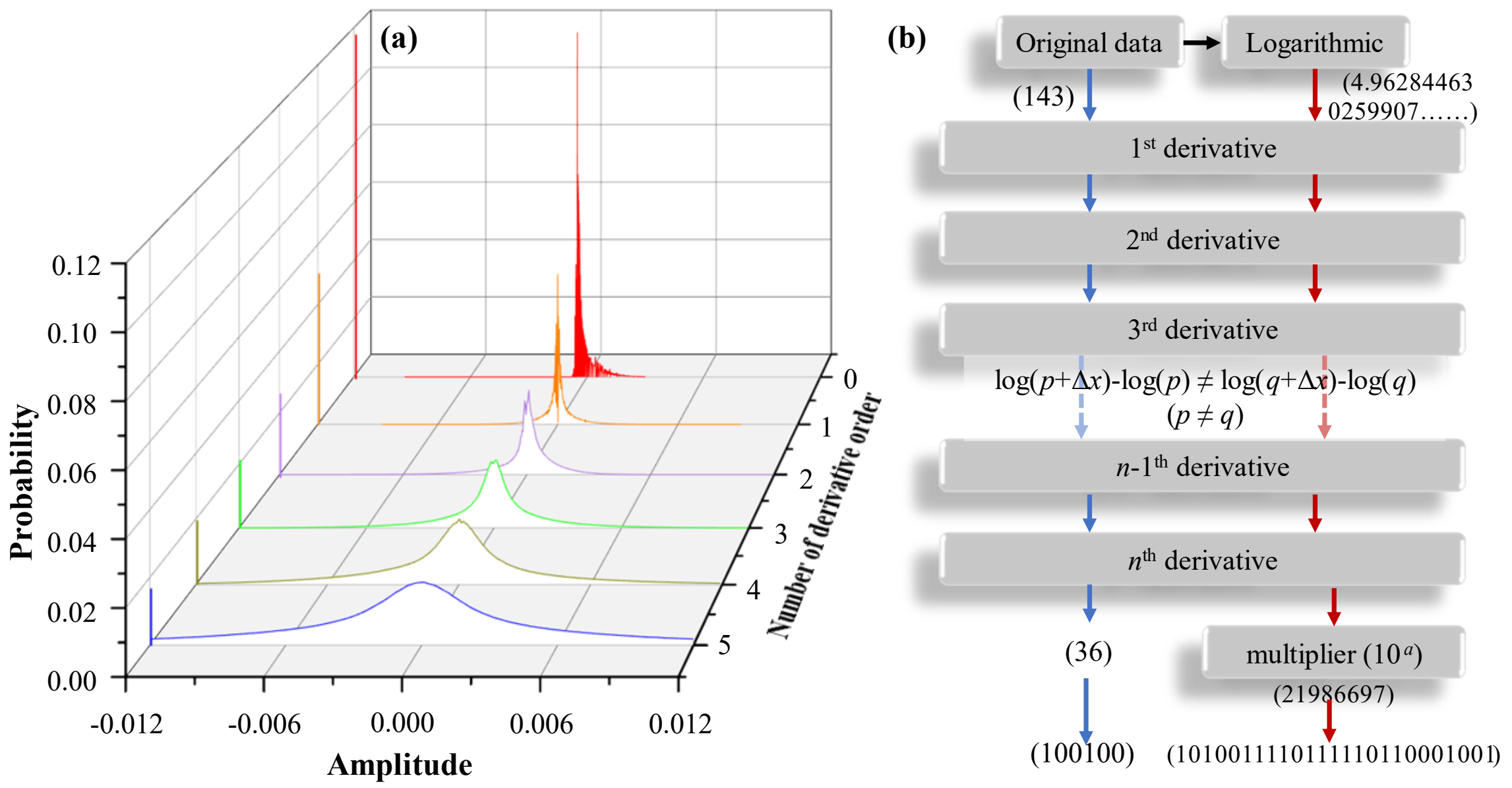

Fig. 7 | (a) Amplitude distribution of derivative chaotic signal of the recorded 2nd-order Stokes light. (b) Schematic of the logarithmic process and high-order derivative process in improving the randomness and the bit rate.

Fig. 8(a) shows the evolution of the standard deviation of the statistical counts of the amplitude converted from the retained LSBs of the 2nd-order Stokes light processed by logarithmic 5th-derivative as a function of the number of retained LSBs. It can be seen that as the number of retained LSB reduces, the standard deviation of the counts gradually decreases, which indicates that the proportions of the generated random bit sequence which contains different bit patterns gradually become uniform. The specific amplitude distributions are also shown in the insets of Fig. 8(a). The amplitudes in the horizontal coordinate are the decimal numbers converted from the 33-LSBs, 23-LSBs, 17-LSBs and 14-LSBs of the digitalized random amplitude, respectively. For example, the decimal integers in the horizontal coordinate of Fig. 8(a1) are converted from the corresponding 33-bit binary numbers. For the case of 33-LSBs, the distribution histogram shown in Fig. 8(a1) is similar to the above-mentioned Gauss-like distribution indicated by the blue solid curve in Fig. 7(a), which deteriorates the randomness if the random numbers are extracted directly from these waveforms. For the case of 23-LSBs, the histogram also shows a non-uniform distribution as illustrated in Fig. 8(a2). As the number of retained LSBs is further decreased, the uniformity of the histogram distribution is improved as shown in Fig. 8(a3). For the case of 14-LSBs, the distribution histogram shown in Fig. 8(a4) are almost uniform, which can be regarded as prerequisite of verified randomness. The uniform distribution indicates that the generated bit sequence contains different bit patterns with equal probabilities. To determine the maximum number of retained LSBs for random bit generation, the NIST test suite with 15 statistical test items is performed on the random bit sequences generated with different numbers of retained LSBs according to the test rules described in Methods: NIST test. Fig. 8(b) shows the number of passed NIST test items for random bit sequences generated from the 31 channels as a function of the number of retained LSBs. The vertical axis denotes the number of the NIST test items that is passed by the generated random bit sequence, where “15” corresponds to the results that all the 15 NIST test items are passed. It can be seen from Fig. 8(b) that the maximum retained LSBs with verified randomness is 14 and the corresponding random bit generation rate at the sampling rate of 2.5 GS/s is 14 × 2.5 GS/s = 35 Gbps. However, it is worth noting that this bit rate can also be boosted by employing a higher sampling rate.

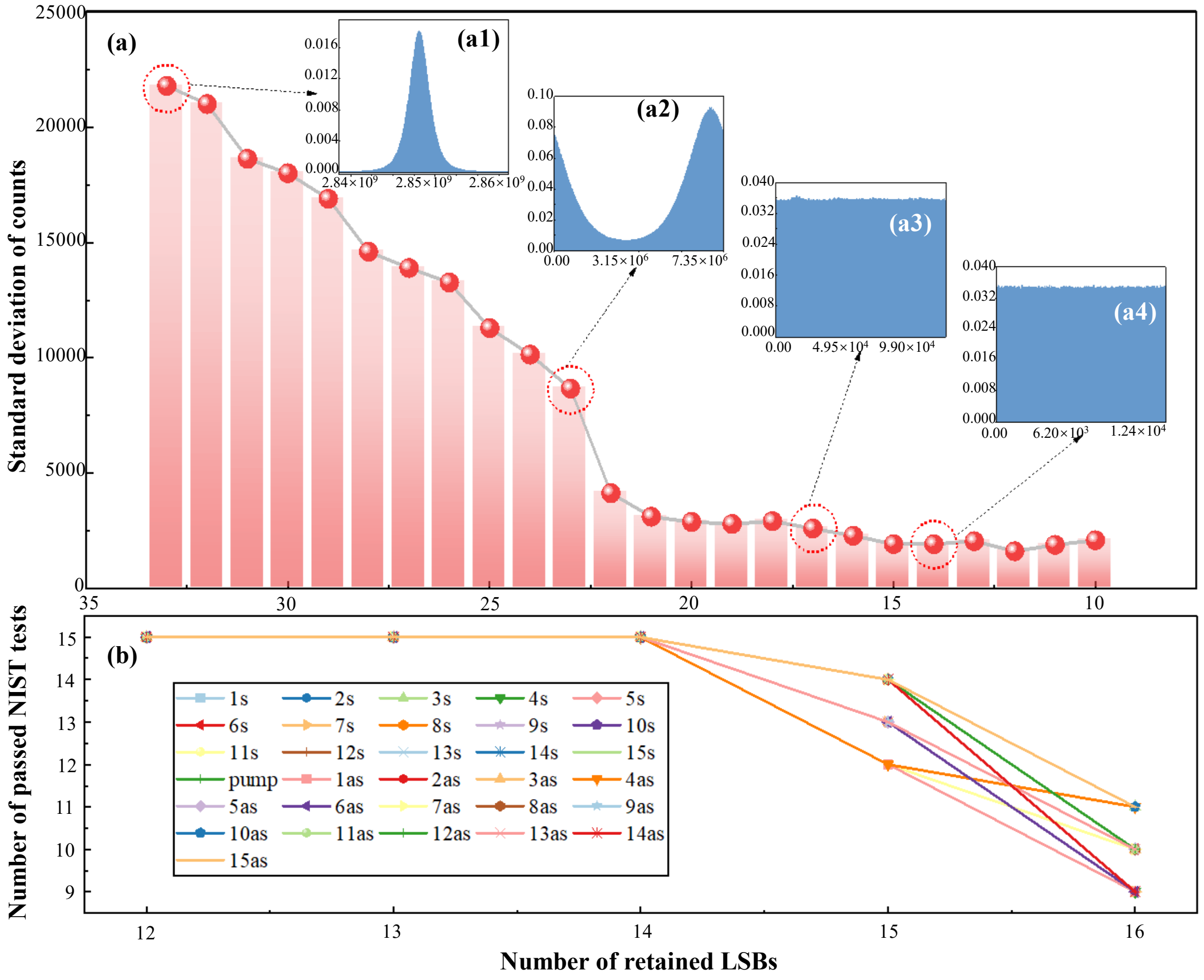

Fig. 8 | (a) Evolution of the standard deviation of the counts of the amplitude distributions of digitalized 2nd-order Stokes light by logarithmic 5th-derivation with different numbers of retained LSBs and the specific amplitude distribution with the retained 33 LSBs (a1), 23 LSBs (a2), 17 LSBs (a3) and 14 LSBs (a4); (b) Number of passed NIST tests as a function of the number of retained LSBs.

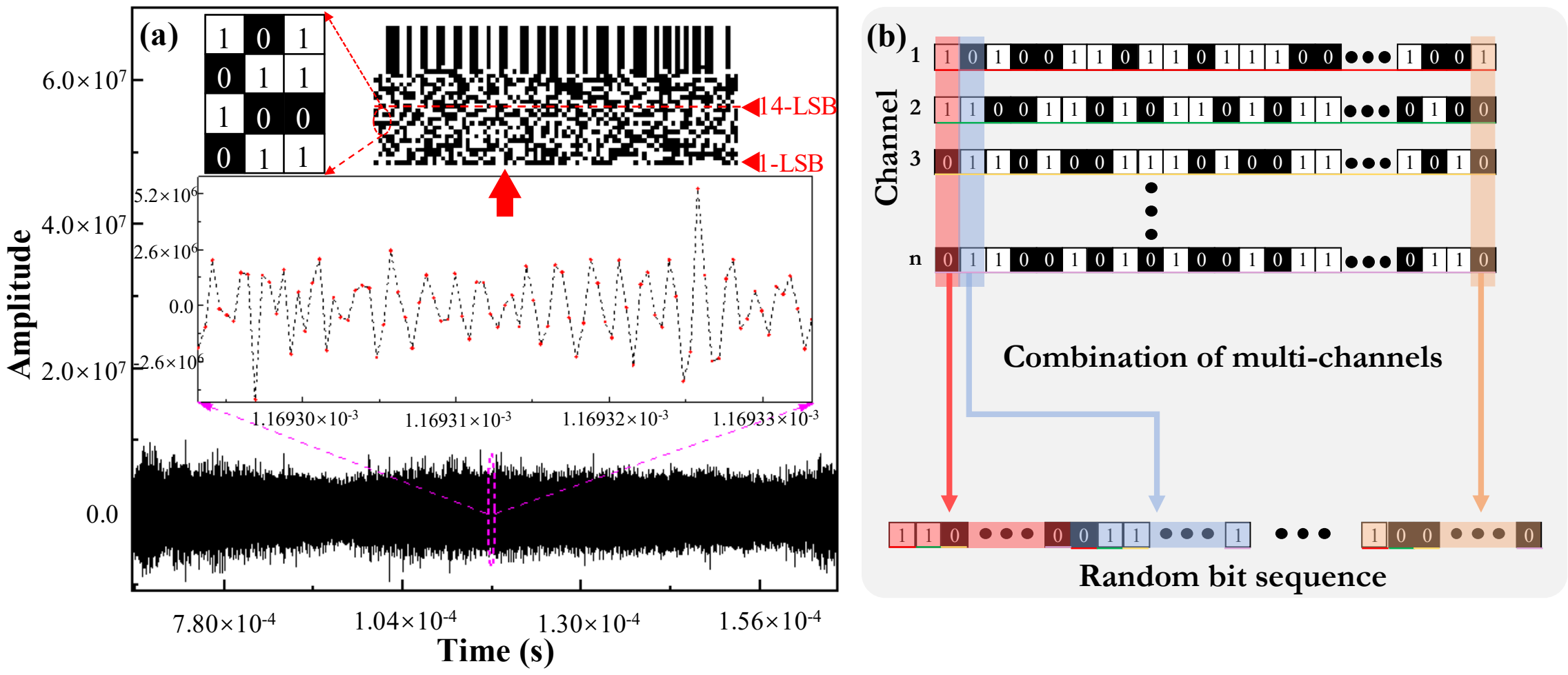

Fig. 9 | Schematic of the random bit generation procedure and examples of digitized random bit data (a), and the bitwise interleaving operation on binary streams from multiple channels (b).

Subsequently, the chaotic signals processed by logarithmic-based high-order-derivation are transformed into binary sequences of “0 1” values to generate a random number stream. A section of the chaotic temporal trace of the 1$^{st}$-order Stokes light acquired at a sampling rate of 2.5 GS/s and processed with the logarithmic-based 5$^{th}$-order derivative is presented in Fig. 9(a). From the close-up view of this signal, it is intuitively clear that this signal exhibits distinct temporal chaotic behavior. The chaotic signal in the close-up region is then directly converted to binary “0 1” values, where the 33-bit binary series generated by each sample point is distributed in time order as shown in the inset in the upper right of Fig. 9(a). The black squares indicate the value “0” and the white squares indicate the value “1” as illustrated in the inset in the upper left of Fig. 9(a). The randomness of this binary number sequence can be clearly observed from its two-dimensional distribution inset. Note that with high-order derivatives and multiplier, the substantial derivative values, reaching up to $10^6$, would display consecutive sequences of "0" or "1" values for most significant bits (MSBs) of the converted binary values, in which high randomness cannot be guaranteed when a large number of LSBs are retained for appending the random bit streams. Therefore, to ensure the randomness of the generated random bit sequences, the number of retained LSBs should be selected carefully to generate high-quality random bit sequences. As mentioned above, here 14-LSBs is retained for random bit sequence owing to the fact that it is the highest LSB that can be optionally retained to guarantee the optimum randomness of the generated random bits sequence and meanwhile ensure the bit generation rate as high as possible. Moreover, since random bit streams can be generated simultaneously in the 31 channels of the STU-BRFL, multiplication of the random bit generation rate can be realized by using the bitwise interleaving operation on these binary streams thanks to the spectrotemporally uncorrelated feature of the BRFL, as illustrated in Fig. 9(b). By combining random bits from multiple channels, a final random bit sequence with a rate up to 1.085 Tbps (2.5-GS/s × 14-LSBs × 31-channels) is achieved by the current experimental setup.

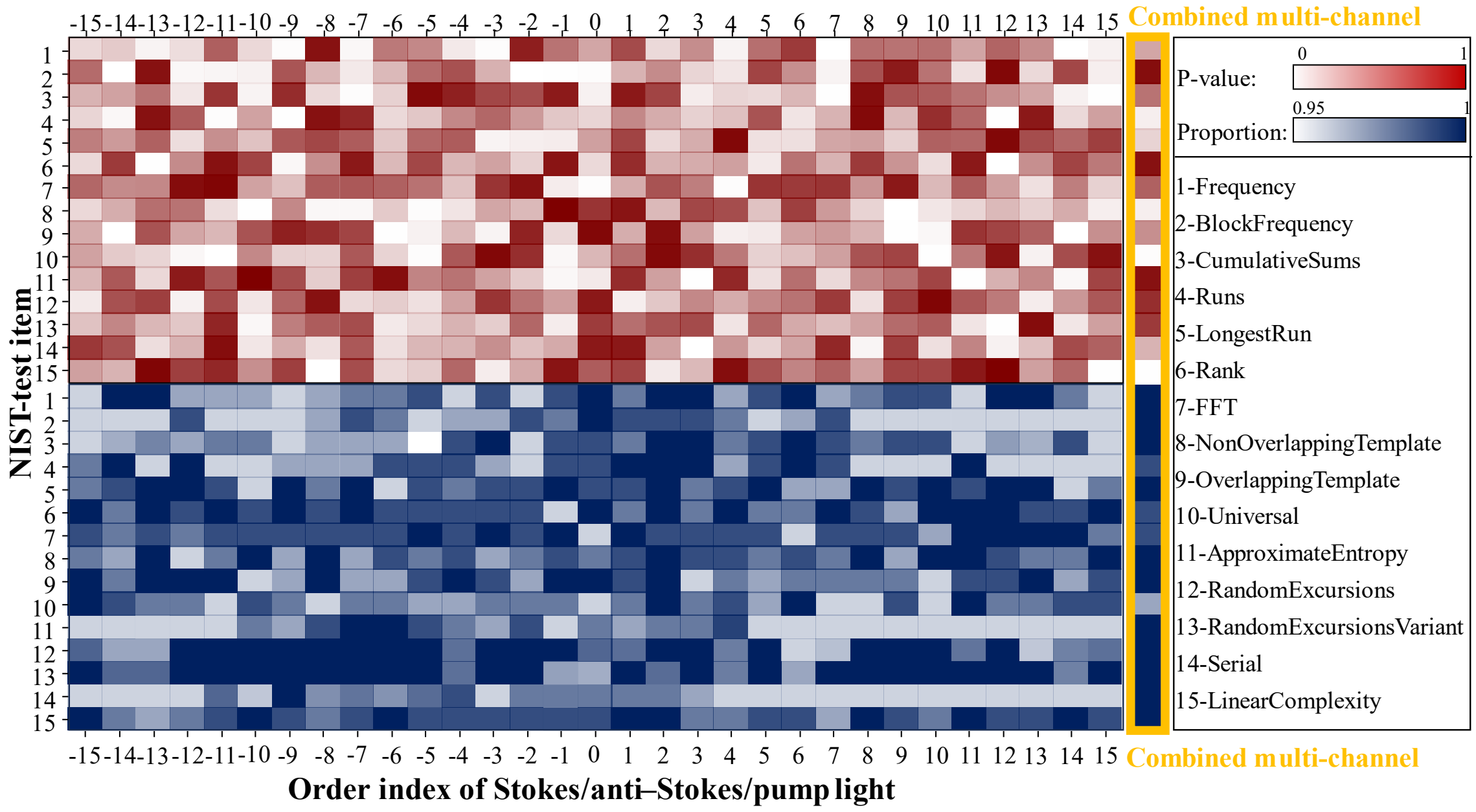

Fig. 10 | NIST test results for random bit sequences generated in 31 single channels and the combined-31-channels by using the first 15 orders of Stokes lights, the first 15 orders of anti-Stokes lights and the residual pump light. (To pass the NIST test, the P-value should be larger than $1\times10^{-4}$ and the corresponding proportion should be greater than 0.96.)

A strict randomness test of the generated random bits is carried out using the NIST test suites. For random bit sequences generated with 14-LSBs retained for each channel, typical NIST tests results for 31 single-channel at the bit rate of 35 Gbps and bitwise-interleaved combination of 31-channels at the bit rate of 1.085 Tbps are presented in Fig. 10. All the generated random bit sequences from single- and combined-multi-channel successfully pass the NIST test with P-values >0.0001 and pass rates >96%. In addition, statistical analysis of bit “0” and “1” for the generated random bit sequences are also performed in the Supplementary Note III.

## Conclusions

In summary, a high-quality chaotic light source with multiple parallel channels based on a spectrotemporally uncorrelated multi-wavelength Brillouin random lasing oscillation has been proposed and demonstrated, which is successfully implemented for parallel fast random bit generation. The inherent correlation among multiple Stokes/anti-Stokes lights is effectively released by the chaotic instabilities induced by the random distribution feedback and intense mode competition which are enhanced and disorderly redistributed among different lasing lines through the complex nonlinear interactions, realizing a chaotic laser source with multiple spectrotemporally uncorrelated channels and making parallel random bit generator a reality. A logarithm-based high-order derivative data processing method is proposed to eliminate the bias and improve the randomness of chaotic signals, increase the available number of LSBs and eventually significantly boost the rate of random bit generation. In the experiment, parallel random bit generation is demonstrated with 31 channels, 35-Gbps single-channel bit rate and 1.085-Tbps total bit rate using 2.5-GS/s sampling rate, $5^{th}$ derivative and 14 LSBs. NIST statistic tests verify the randomness of the bit streams generated in each single-channel and the combined-multi-channel.

## Methods

### Experimental details of the STU-BRFL

Fig. 2 shows the experimental setup of the STU-BRFL-based PRBG, which consists of a main cavity for random lasing oscillation and a sub-cavity for light amplification and re-injection of light into the main cavity. A 390-m-long highly nonlinear fiber with nonlinear coefficient of 10 $W^{-1}\cdot km^{-1}$, dispersion of 0.5753183 $ps\cdot nm^{-1}\cdot km^{-1}$ (@1550nm) and zero-dispersion wavelength of 1521.78962 nm is used for providing Brillouin gain and parametric gain. A 5-km-long conventional single-mode fiber is used for providing random distributed feedback. The laser is pumped by a 1550 nm semiconductor laser with a linewidth of ~15 kHz. In the main cavity, the pump light is first launched into the Brillouin gain fiber. When the pump power is increased to surpass the SBS threshold of the Brillouin gain fiber, the SBS process would be triggered through electrostriction, giving rise to the amplified counter-propagating Stokes light. The generated Stokes light passes CIR1 and is then equally divided by OC3 into two parts, with one part entering the sub-cavity and the other part continuing propagating in the main cavity. The part of Stokes light remained in the main cavity then enters the 5-km-long SMF through PC3 and CIR2. Within the 5-km-long SMF, the Stokes light is randomly scattered along the fiber through Rayleigh scattering and then sent back to the Brillouin gain fiber through CIR2 as random distributed feedback to rejoin the Brillouin amplification process for lasing resonances. The Stokes light is repeatedly amplified each time it circulates in the main cavity, which results in a rapid increase in its power. The other part of the Stokes light is guided into the sub-cavity by OC3 and mixed with the pump light through OC1 after power adjustment by VOA1. After being amplified by the EDFA, the mixed light is re-injected into the main cavity through CIR1 as the Brillouin pump light to generate higher-order Stokes lights through the CSBS process. Meanwhile, the quasi-phase-matched FWM is also excited among the Stokes/anti-Stokes lights and the pump light to further boost the power of existing Stokes lights and generate higher-order Stokes and anti-Stokes lights. As the gradually accumulated effective optical nonlinear gain for the Stokes/anti-Stokes lights exceeds the total loss in the entire cavity, the circulating Stokes/anti-Stokes lights would start to resonate in the main cavity and eventually build up the multi-wavelength random lasing oscillation. It is worth noting that the NOLM is added in the main cavity to function as a power balancer to decrease the peak power discrepancy among multiple Stokes/anti-Stokes lights. As mentioned above, for the resonant Stokes/anti-Stokes lights of each order, the random distributed feedback provided by the Rayleigh scattering and the intense mode competition for finite Brillouin gain result in the chaotic instabilities of the lasing output. Moreover, these chaotic instabilities are intensified and disorderly redistributed among multiple lasing lines through the complex nonlinear interactions between the CSBS process, the quasi-phase-matched FWM process and the random mode resonance process, leading to the effective removal of the inherent correlation between multiple Stokes and anti-Stokes lights. Finally, the spectrotemporally uncorrelated multi-wavelength random lasing is successfully built up and emitted at the output port. To realize PRBG, the generated random lasing output is demultiplexed into multiple single-wavelength Stokes/anti-Stokes lines of different orders by a customized WD. Temporal output of each demultiplexed Stokes/anti-Stokes line is then acquired by a PD, which converts the optical signal into the electrical signal. After data processing, high-rate random bit sequences are simultaneously generated in multiple channels.

### Characterization of random laser output

After data recording of the demultiplexed Stokes/anti-Stokes lights as well as the residual pump light, the intra-channel and inter-channel correlations are calculated by the autocorrelation function ACF and the cross-correlation function XCF

$$C_{\mathrm{ACF}}(\tau)=\frac{\left\langle\left(I_a(t)-\langle I_a(t)\rangle\right)\left(I_a(t+\tau)-\langle I_a(t)\rangle\right)\right\rangle}{\left\{\left\langle\left(I_a(t)-\langle I_a(t)\rangle\right)^2\right\rangle\left\langle\left(I_a(t+\tau)-\langle I_a(t)\rangle\right)^2\right\rangle\right\}^{1/2}} \tag{1}$$

$$C_{\mathrm{XCF}}(\tau)=\frac{\left\langle\left(I_a(t)-\langle I_a(t)\rangle\right)\left(I_b(t+\tau)-\langle I_b(t)\rangle\right)\right\rangle}{\left\{\left\langle\left(I_a(t)-\langle I_a(t)\rangle\right)^2\right\rangle\left\langle\left(I_b(t+\tau)-\langle I_b(t)\rangle\right)^2\right\rangle\right\}^{1/2}} \quad (2)$$

where $I_a(t)$ and $I_b(t)$ are light intensities recorded in channel $a$ and $b$, $\tau$ is the delay time, and the $< >$ denotes the time average.

### NIST test

The state-of-the-art National Institute of Standards and Technology (NIST Special Publication 800-22) test suite with 15 statistical test items is used to verify the statistical randomness of the generated random bit sequences in 31 parallel channels. As advised by the NIST, each test is performed using 100×1 Mbits with a statistical significance level $\alpha$ = 0.01[32]. The test criterion for "success" is that the P-value (uniformity of p-values) should be larger than 0.0001 and the proportion should be larger than 0.96[31]. Note that the $P$-value is the probability that a perfect random bit generation would have produced a sequence less random than the sequence that was tested, given the kind of non-randomness assessed by NIST tests.

### Random bit generation

The digital post-processing procedure for random bit generation is illustrated in Fig. 7(b). A chaotic signal value of 143, which falls in the range of 0 ~ 4096 and recorded by a 12-bit ADC, is used as an example to describe the processing procedure. First, logarithmic processing is performed on this value, which is converted to a value (e.g., 4.962844630259907…) with multiple decimal places. In the subsequent high-order derivative process, the randomness of these logarithmic chaotic signals is improved compared to the original raw data, mainly due to the fact that $\log(p+\Delta x) - \log(p) \neq \log(q+\Delta x) - \log(q)$ ($p \neq q$). Eventually, by moving the decimal point of the $n^{\mathrm{th}}$ derivative chaotic value to the right through the multiplier, one can obtain a larger decimal number (e.g., 21986697) that is converted to binary (e.g., 1010011110111110110001001) after rounding. It should be noted that the value of $a$ for the multiplier used in the present experiment is 8, which is the maximum value preferred without generating additional invalid MSBs. It is clear that after the above operation there are more LSBs that can be chosen to be appended to the random bit sequences.

## Acknowledgements

This works was supported by the National Natural Science Foundation of China (62105180), Natural Science Foundation of Shandong Province (ZR2020MF110, ZR2020MF118), Taishan Scholar Foundation of Shandong Province (tsqn202211027), National Grant Program for High-level Returning Oversea Talents (2023), Qilu Young Scholar Program of Shandong University and Shandong Higher School Youth Innovation Team Technology Program (2022KJ024).

## Author contributions

Conceptualization, Yanping Xu.; methodology, Yanping Xu; software, Yuxi Pang; validation, Yanping Xu and Yuxi Pang; formal analysis, Yanping Xu and Yuxi Pang; investigation, Yanping Xu, Yuxi Pang, Shaonian Ma and Qiang Ji; resources, Yanping Xu; data curation, Yuxi Pang, Qiang Ji and Yanping Xu; writing—original draft preparation, Yanping Xu and Yuxi Pang; writing—review and editing, Yanping Xu, Ping Lu and Xiaoyi Bao; visualization, Yuxi Pang and Yanping Xu; supervision, Yanping Xu, Xian Zhao, Zengguang Qin, Zhaojun Liu, Ping Lu and Xiaoyi Bao; project administration, Yanping Xu; funding acquisition, Yanping Xu. All authors have read and agreed to the published version of the manuscript.

## Competing interests

The authors declare no competing financial interests.

# Authors:

**Yanping Xu** received the B.S. degree in physics from Jilin University, Changchun, China, in June 2011, and the Ph.D. degree in physics from the University of Ottawa, Ottawa, ON, Canada, in August 2017. He is currently a Full Professor with the Center for Optics Research and Engineering, Shandong University, Qingdao, China and has been conferred the title of Taishan Young Scholar of Shandong Province and Qilu Young Scholar of Shandong University. From 2017 to 2018, he was with the Department of Physics, University of Ottawa, as a Postdoctoral Fellow. In 2018, he joined Ciena Corporation (Originally known as Nortel Networks Corporation), Ottawa, ON, Canada, as a Fiber-Optic Communication Research Scientist. He has authored or co-authored more than 60 research papers in peer-reviewed journals and conference proceedings. His research interests include fiber-optic sensors, fiber lasers, random number generation, nonlinear fiber-optic effects, and fiber-optic communications. He has been invited to give oral presentations in OFS, OFC, CLEO, IEEE Sensors, and Photonics North conferences. He was also the Guest Editor of the Sensors and a Reviewer of peer-reviewed journals, including the Photonics Research, Optics Express, Optics Letters, IEEE Journal of Lightwave Technology, IEEE Photonics Technology Letters, IEEE Photonics Journal, Applied Optics, Optics Communications, and Optical Fiber Technology.

**Yuxi Pang** received his BS degree from Shandong Jianzhu University in 2020, and his MS degree from Shandong University in 2023. He is currently working toward the PhD degree in optical engineering with the Center for Optics Research and Engineering and Key Laboratory of Laser and Infrared System of the Ministry of Education, Shandong University, Qingdao, China. His research interests include random fiber laser, fiber laser, and optical fiber sensing.

Supplementary Information for

# Parallel fast random bit generation based on spectrotemporally uncorrelated Brillouin random fiber lasing oscillation

**Yuxi Pang,[1,2] Shaonian Ma,[1,2] Qiang Ji,[1,2] Xian Zhao,[1,2] Zengguang Qin,[2,3] Zhaojun Liu,[2,3] Ping Lu,[4] Xiaoyi Bao,[5] and Yanping Xu,[1,2,*]**

[1]*Center for Optics Research and Engineering, Shandong University, Qingdao 266237, China*

[2]*Key Laboratory of Laser and Infrared System of the Ministry of Education, Shandong University, Qingdao 266237, China*

[3]*School of Information Science and Engineering, Shandong University, Qingdao 266237, China*

[4]*National Research Council Canada, 100 Sussex Drive, Ottawa, ON K1A 0R6, Canada*

[5] *Physics Department, University of Ottawa, 25 Templeton Street, Ottawa, ON K1N 6N5, Canada*

**Corresponding author: yanpingxu@sdu.edu.cn*

## Supplementary Note I: Spectral properties of the spectrotemporally uncorrelated Brillouin random fiber laser

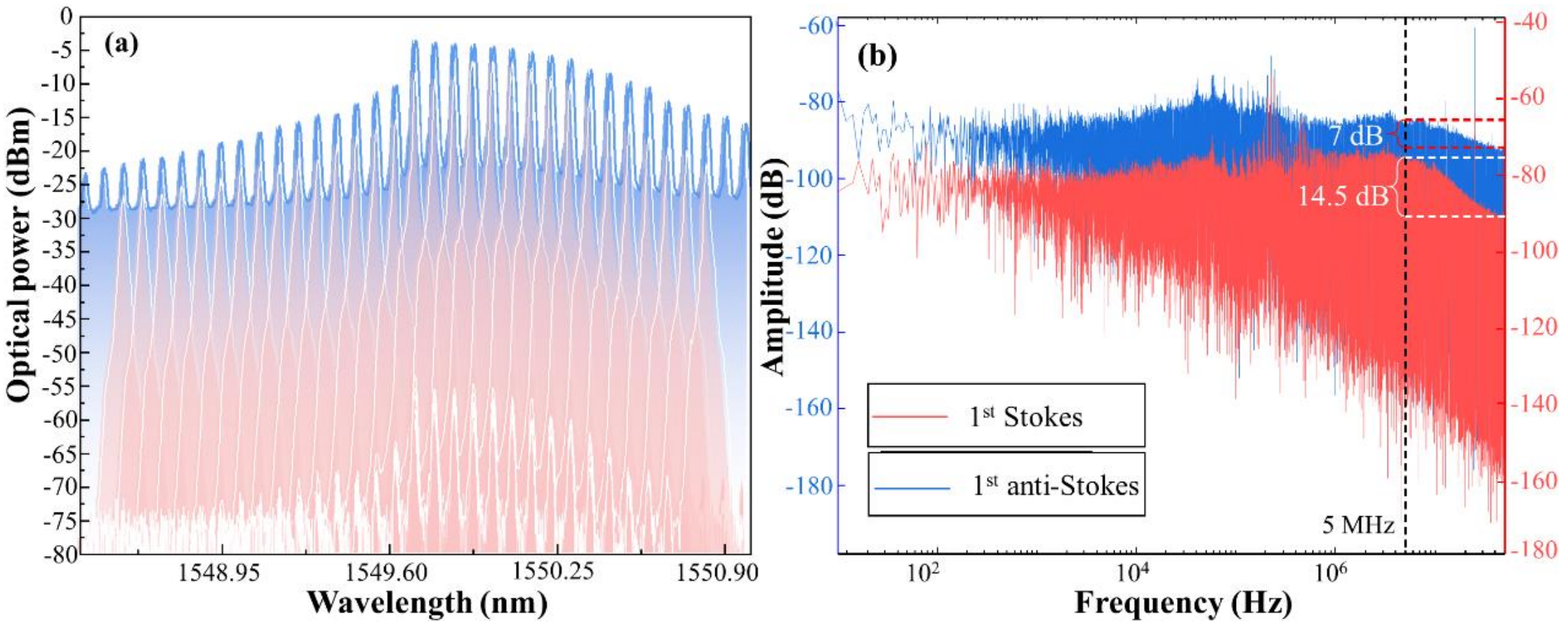

**Supplementary Fig. 1** (a) Compilation of the optical spectra of the demultiplexed Stokes and anti-Stokes lights of each order as well as the residual pump light; (b) Power spectra of 1st-order Stokes and 1st-order anti-Stokes lights.

The random lasing emission of the STU-BRFL is demultiplexed by a wavelength demultiplexer and the demultiplexed Stokes/anti-Stokes lines are then individually recorded for optical and electrical spectrum analysis. Fig. 1(a) shows a compilation of the optical spectra of the demultiplexed Stokes and anti-Stokes light of each order as well as the residual pump light, in which high OSNRs are guaranteed for each demultiplexed channel. In Fig. 1(b), power spectra of the 1st-order Stokes light (red curve) and the 1st-order anti-Stokes light (blue curve) exhibit a relatively flat plateau in the frequency range from 10 Hz to several megahertz, corresponding to chaotic bandwidths in an order of MHz for both channels. The falling edge of the power spectra is associated with the bandwidth of the Brillouin gain spectrum and determined by the phonon lifetime (several tens of nanoseconds) in the single-mode fiber (SMF), which limits the maximum bit rate that can be achieved by directly utilizing the random lasing output. By comparing the two power spectra, it is found that the amplitude drops in the range from the falling edge of ~5 MHz to the maximum monitored frequency of 50 MHz is decreased from ~14.5 dB for the 1st-order Stokes light to ~7 dB for the 1st-order anti-Stokes light. This result shows that the chaotic bandwidth of the anti-Stokes random lasing output is somehow larger than that of the Stokes random lasing output, owing to the more pronounced chaotic instabilities occurring in the anti-Stokes resonances since the quasi-phase-matched FWM process and random mode resonance process that bring noticeable instabilities play a dominant role in the lasing process of the anti-Stokes lights.

**Supplementary Note II: Correlation analysis of the spectrotemporally uncorrelated Brillouin random fiber laser**

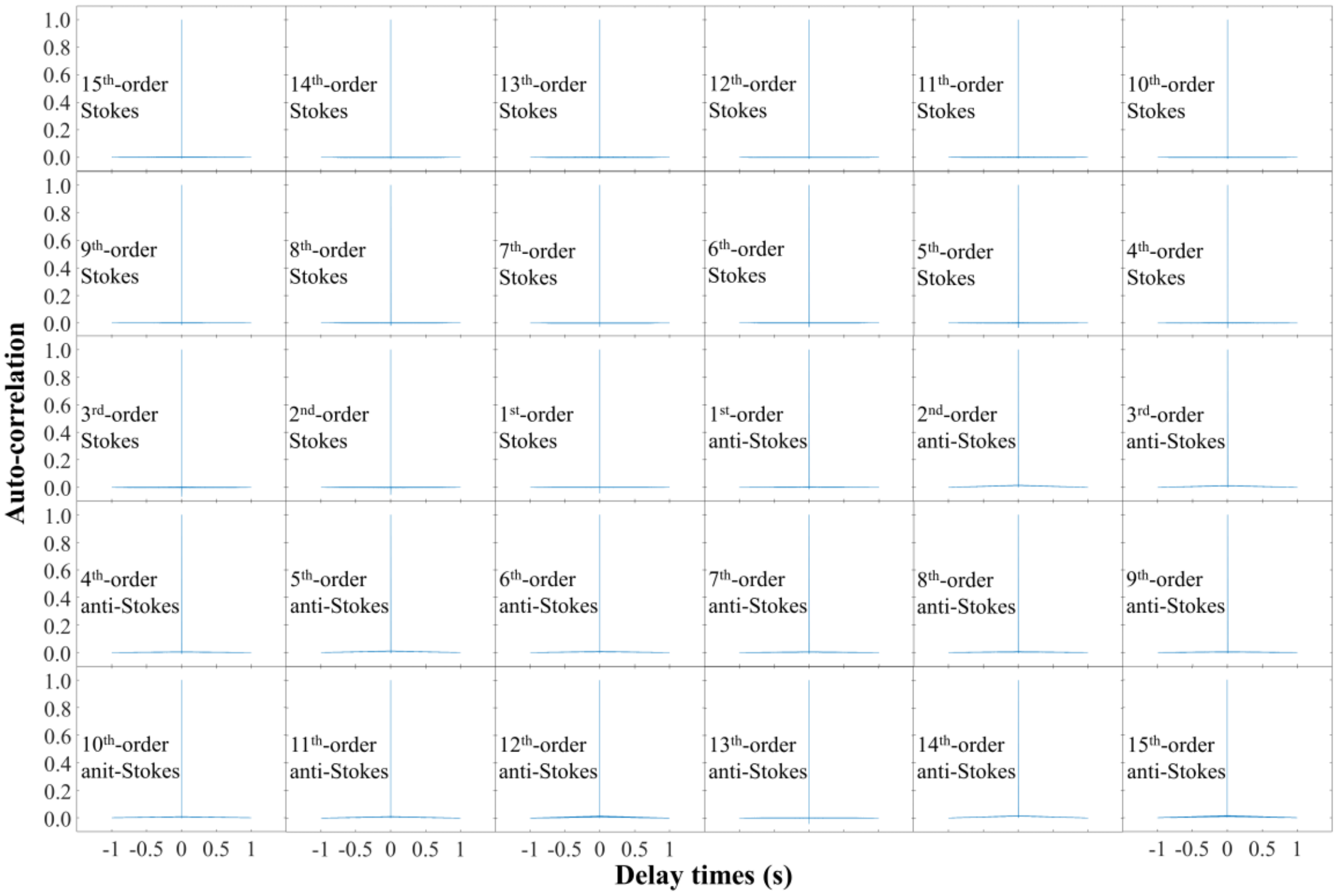

**Supplementary Fig. 2** Auto-correlation functions of time traces acquired from all the 15 orders of Stokes lights and 15 orders of anti-Stokes lights.

The intra-channel correlation results calculated by auto-correlation function (ACF) for all the 15 orders of Stokes lights and 15 orders of anti-Stokes lights as well as the residual pump light are shown in Fig. 2. Obviously, all the ACF results present a $\delta$-function shape without any side peaks, indicating that no time delay signature exists in the random lasing output. Thus, unpredictability and randomness are ensured, demonstrating that each lasing line of the STU-BRFL can individually function as an effective chaotic light source.

In addition, the inter-channel correlation between the demultiplexed light components in the STU-BRFL is calculated by cross-correlation function (XCF). The calculated XCF results between all 15 orders of Stokes lights, 15 orders of anti-Stokes lights and the residual pump light are shown in Fig. 3. It can be found that none of the cross-correlation results exhibit any significant correlation peak and the calculated XCF coefficient level maintains near zero, which verifies the independence among the selected channels of the STU-BRFL output. Furthermore, it can be clearly observed that the correlation calculations located on the diagonal of Fig. 3 all exhibit extremely high values at zero time delay, which is actually the autocorrelation function corresponding to the different laser lines.

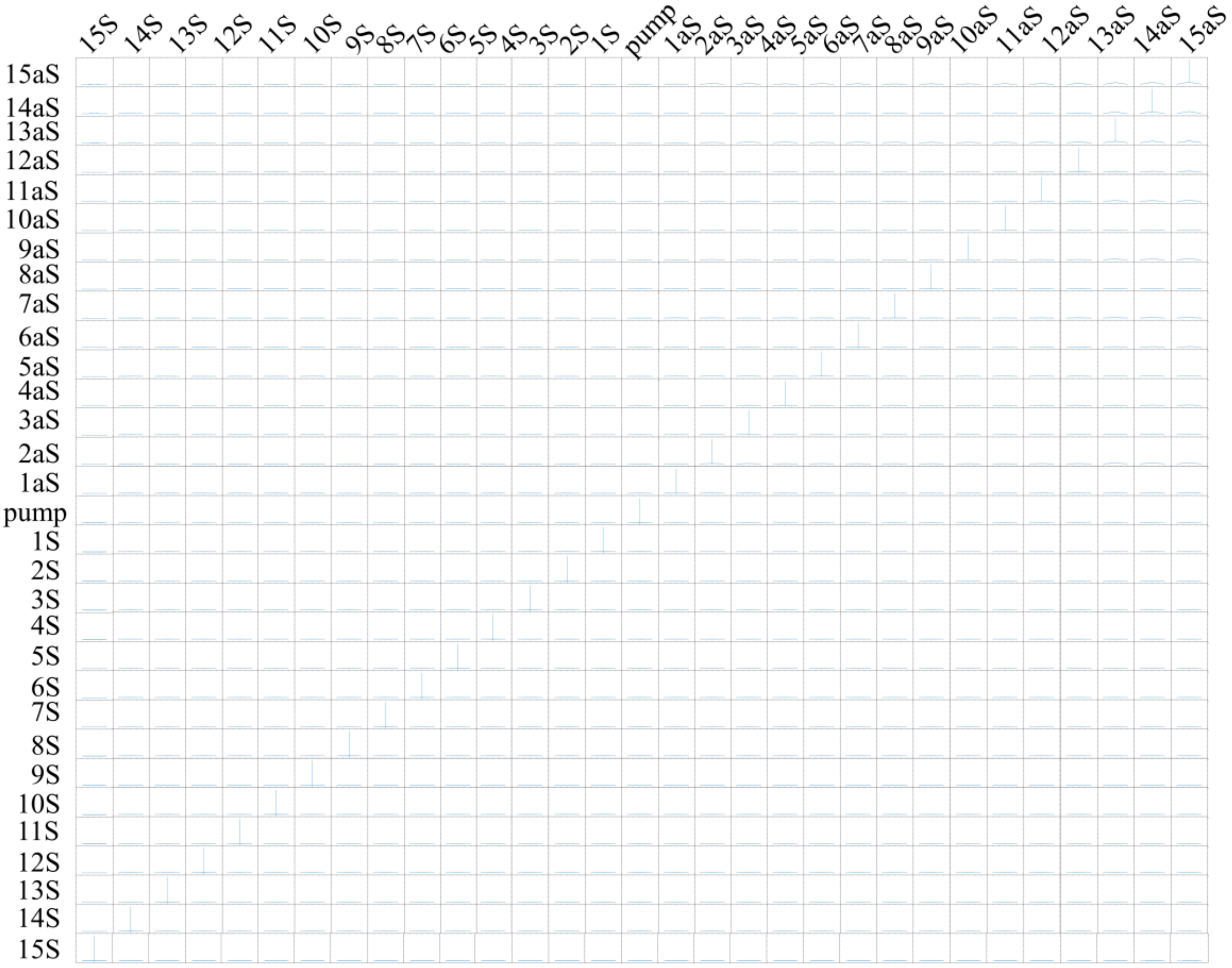

**Supplementary Fig. 3** Cross-correlation function of time traces between all the 15 orders of Stokes lights, 15 orders of anti-Stokes lights and the residual pump light.

## Supplementary Note Ⅲ: Characterization of the randomness of the generated random bit sequences

Fig. 4(a) shows the statistical results of the generated random bit sequence at the rate of 35 Gbps in a single RBG channel using temporal output of the $1^{st}$-order Stokes light. The statistics of bit “0” and “1” verify the two bins of 0 and 1 are of nearly equal size in the generated random bit sequences, as shown in Fig. 4(a1). The ratio of bit “0” is calculated as 0.500029. An example bit map with blue and red squares constructed from 500×500 bits is depicted in Fig. 4(a2). As could be expected from the random bit generation process, no obvious pattern or bias is observed in the bit map. The statistical distribution and the bit map of the combined-31-channels random bit sequence are shown in Fig. 4(b1~b2), which further demonstrates the multi-channel multiplexing capability of this PRBG for achieving higher-rate and truly random number generation.

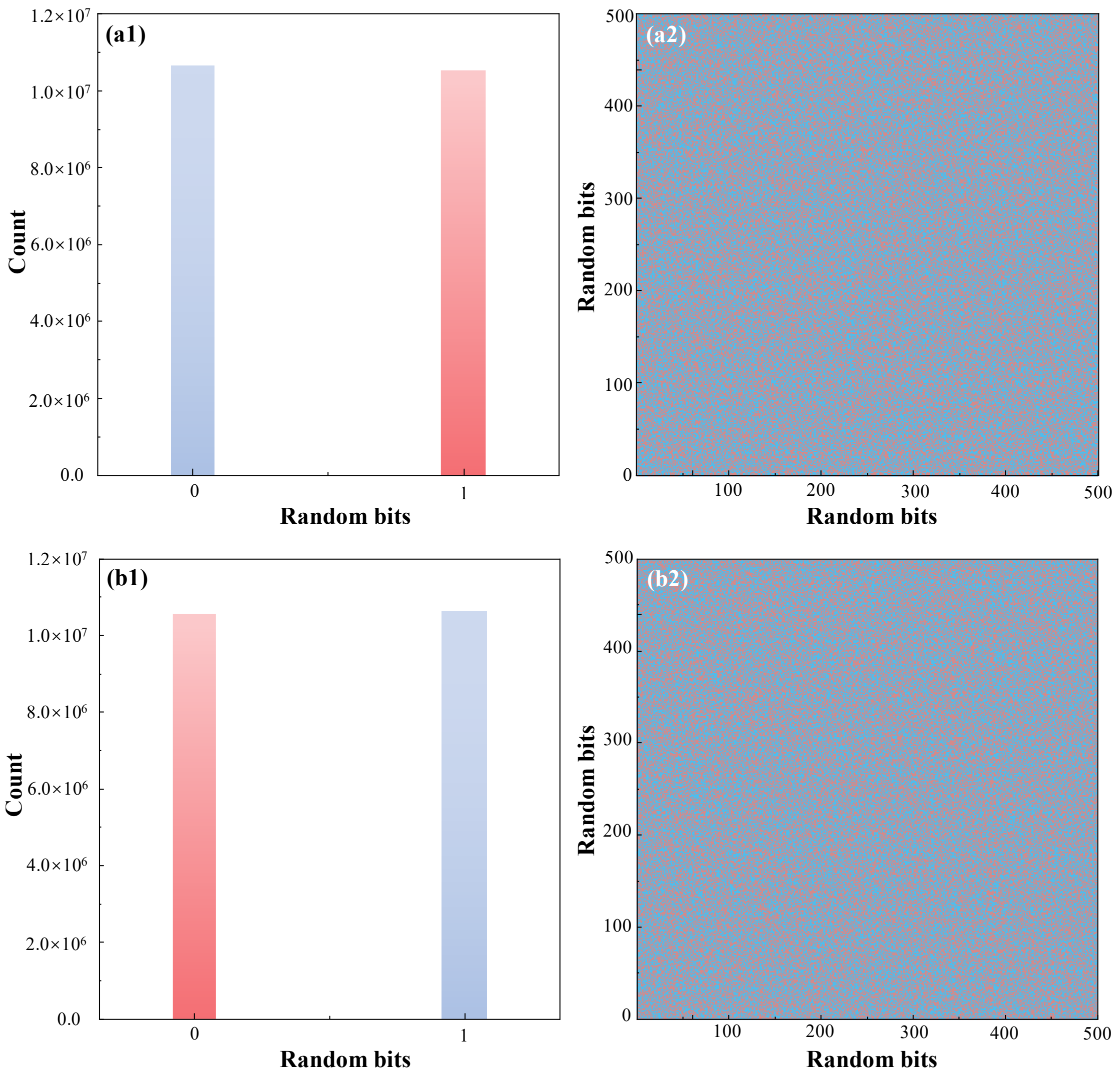

**Supplementary Fig. 4** Statistics of 0 and 1 for random bit streams generated with a single channel using temporal output of the 1st-order Stokes light (a1) and the combined 31 channels (b1), respectively. Examples of bit map plotted in a two-dimensional plane using 500×500 random bits generated with the single channel (a2) and the combined 31 channels (b2), respectively.